\documentclass[graphics,floatfix, nofootinbib,tightenlines,nobibnotes, 12pt]{revtex4-2}
\usepackage{hyperref}
\hypersetup{hypertexnames=false}
\usepackage{graphicx}
\usepackage[english]{babel}
\usepackage[utf8]{inputenc}
\usepackage{amsmath,amssymb}
\usepackage{amsfonts}
\usepackage{multirow}
\usepackage{float}
\usepackage{subfigure}
\usepackage{color}
\graphicspath{{fig/}}
\usepackage{slashed}
\usepackage{array}[=2016-10-06]
\usepackage{diagbox}
\usepackage{tabularray}
\usepackage{tabularx}
\usepackage[normalem]{ulem}
\usepackage{cancel}

\begin{document}

\title{Investigation of the photoproduction of bottom-charmed baryons within NRQCD}

\author{Juan-Juan Niu$^{1}$}
\email[Contact author: ]{niujj@gxnu.edu.cn}
\author{Hong-Hao Ma$^{2,1}$}
\email[Contact author: ]{honghao.ma@unesp.br}

\address{$^{1}$ Department of Physics, Guangxi Normal University,\\Guilin 541004, People's Republic of China}
\address{$^{2}$ Instituto de F$\acute{\imath}$sica Te$\acute{o}$rica, Universidade Estadual Paulista,\\
Rua Dr. Bento Teobaldo Ferraz, 271-Bloco II, 01140-070 S$\tilde{a}$o Paulo, S$\tilde{a}$o Paulo, Brazil}

\date[]{}
\received[Received ]{2 February 2026}
\revised[revised ]{21 July 2026}
\accepted[accepted ]{4 September 2026}
\published[published ]{29 September 2026}

\begin{abstract}

We present a further theoretical study of the orbital $P$-wave bottom-charmed baryon within the framework of nonrelativistic QCD, considering both the direct photoproduction channel $\gamma+\gamma \rightarrow \Xi_{bc} +\bar{c}+\bar{b}$ and the resolved photoproduction channel $\gamma+g \rightarrow \Xi_{bc} +\bar{c}+\bar{b}$. At future linear colliders, ILC and CLIC, the initial photons can be emitted from the laser backscattering and then the parton gluon can be emitted from the photon. The formation of $\Xi_{bc}$ can be modeled in two steps: a compact diquark state $\langle bc\rangle[n]$ is formed first and subsequently captures a light quark from the vacuum to hadronize into the baryon $\Xi_{bc}$. The color and spin quantum number $[n]$ of the $\langle bc\rangle$ diquark can be $[{}^3S_1]_{\bar{\textbf{3}}/\textbf{6}}$, $[{}^1S_0]_{\bar{\textbf{3}}/\textbf{6}}$, $[{}^1P_1]_{\bar{\textbf{3}}/\textbf{6}}$, or $[{}^3P_J]_{\bar{\textbf{3}}/\textbf{6}}$
with $J=0,1,2$. Using three benchmark collision energies and an illustrative integrated luminosity, we analyze the cross sections, differential distributions, and the estimated produced events of the $\Xi_{bc}$ baryon. For the conservative $h_{\mathbf{6}}=0$ baseline, the combined $P$-wave contribution amounts to approximately 8\% of the combined $S$-wave rate at three considered collision energies and is therefore non-negligible.\vspace {5mm}
\end{abstract}

\maketitle


\section{Introduction}
\label{sec:1}
Since the discovery of the doubly charmed baryon $\Xi_{cc}^{++}$ by LHCb Collaboration in 2017 \cite{LHCb:2017iph, LHCb:2018pcs, LHCb:2018zpl},
the quark model \cite{GellMann:1964nj,Zweig:1981pd,Zweig:1964jf} has achieved a successful verification, while also presenting new challenges. It predicted the existence of bottom-charmed baryons, which consist of two different heavy quarks ($b$ and $c$) and a light quark ($u$, $d$, or $s$) in multiplets. Because of its heavy mass and the need for both $b$ and $c$ component quarks, compared to the doubly charmed baryons, its cross section is smaller. The LHCb Collaboration using proton-proton collision data performed the first search for the $\Xi_{bc}^{+}(bcu)$, $\Xi_{bc}^{0}(bcd)$, and $\Omega_{bc}^{0}(bcs)$ baryons through the decay channels
$\Xi_{bc}^{+} \rightarrow J/\psi~\Xi_{c}^+$, $\Xi_{bc}^{0} \rightarrow D^0 p K^-$, and $\Xi_{bc}^{0}/\Omega_{bc}^{0} \rightarrow \Xi_{c}^{+}/\Lambda_{c}^{+} \pi^-$ \cite{LHCb:2022fbu,LHCb:2020iko,LHCb:2021xba}. However, these bottom-charmed baryons have not yet had their existence confirmed through experiments to date. Meanwhile, a series of in-depth theoretical studies were carried out using theoretical potential models and phenomenology to focus on the mass, lifetime, production mechanisms, and decay properties of $\Xi_{bc}$ (here, the light quark is not explicitly identified) and their excited states \cite{Gershtein:2000nx,Oudichhya:2022ssc,Qin:2021zqx, Yang:2022nps, Shi:2019hbf,Ghalenovi:2022dok,Mathur:2018epb,Mohajery:2018qhz,Song:2022csw,Song:2023cyk}. The study of these hadrons plays an indispensable role in understanding the quantum chromodynamics (QCD) theory, revealing the interaction between quarks and gluons and exploring the structure of baryons.

With the improvement of the collision energy and luminosity in high-energy physics experiments, the feasibility of discovering the bottom-charmed baryon has been greatly enhanced. At the same time, more precise theoretical calculation results are required for the guidance of experimental exploration.
In the early stages, a large number of theoretical predictions focused on the production mechanisms of doubly heavy baryons $\Xi_{cc}$, $\Xi_{bc}$, and $\Xi_{bb}$ in $S$ waves on multiple platforms, including direct production at hadron colliders \cite{Baranov:1995rc, Chang:2007pp,Chang:2009va,Chen:2014hqa, Chen:2019ykv,Martynenko:2014ola, Koshkarev:2016acq,Koshkarev:2016rci, Groote:2017szb,Berezhnoy:2018bde, Brodsky:2017ntu,Wu:2019gta}, electron-positron colliders \cite{Jiang:2012jt,  Jiang:2013ej,Yang:2014ita, Zheng:2015ixa}, $\gamma\gamma$ \cite{Baranov:1995rc, Yang:2014tca, Chen:2014frw, Zhan:2023jfm} collision, $ep$ collision \cite{Baranov:1995rc, Bi:2017nzv, Sun:2020mvl}, and heavy-ion colliders \cite{Chen:2018koh, Groote:2017szb}, as well as the indirect production through top quark decays~\cite{Niu:2018ycb}, Higgs \cite{Niu:2019xuq}, $W^+$ \cite{Zhang:2022jst}, and $Z^0$ \cite{Luo:2022lcj} boson decay. Subsequently, research on the production mechanism of excited $P$-wave doubly heavy baryons has been carried out successively~\cite{Berezhnoy:1996an, Berezhnoy:2018krl, Zhang:2022jst, Ma:2022cgt, Zhan:2023vwp}. The results show that the contributions of doubly heavy baryons in $P$ waves cannot be negligible, reaching several percent, and it is worthy of further in-depth study. Therefore, at $e^+e^-$ colliders, the photoproduction of doubly heavy baryons $\Xi_{bc}$ in $P$ waves can be conducted.

Regarding the photoproduction of doubly heavy baryons $\Xi_{bc}$ at $e^+e^-$ colliders, it is typically a direct process $\gamma+\gamma \rightarrow \Xi_{bc} +\bar{c}+\bar{b}$, as well as the resolved channel $\gamma +g \rightarrow \Xi_{bc} +\bar{c}+\bar{b}$ \cite{Klasen:2001cu}. The initial $\gamma$ can be emitted either from the bremsstrahlung process and its energy distribution can be well delineated within the Weiz$\ddot{a}$cker-Williams approximation (WWA) \cite{Frixione:1993yw} or from the process of laser backscattering (LBS) of $e^+e^-$ at future linear colliders, ILC and CLIC \cite{ILC:2007bjz, CLICdp:2018cto}. However, the photons from these two sources have different trends. The photon spectrum of LBS shows an upward trend, while that of WWA shows a logarithmic decline trend \cite{Zhan:2023jfm, Jiang:2024lsr}. Future $e^+e^-$ colliders, such as FCC-ee \cite{FCC:2018evy}, CEPC \cite{CEPCStudyGroup:2018rmc, CEPCStudyGroup:2018ghi}, ILC \cite{ILC:2007bjz, Erler:2000jg}, and CLIC \cite{CLICdp:2018cto} will provide a very favorable experimental platform for the search for bottom-charmed baryons. Especially at ILC and CLIC, the contribution of resolved photoproduction channel ($\gamma+g$) of $\Xi_{bc}$ is significant, which is attributed to the LBS photon \cite{Li:2009zzu, Zhan:2022nck, Zhan:2022etq, Ma:2025ito, Niu:2025gcj}. The contribution from a double resolved photoproduction channel is usually very small and can be disregarded. After the LBS photon undergoes a process of resolution, its partons ($\gamma$, $g$, $u$, $d$, $s$) will participate in the subsequent strong interaction, resulting in a high-energy process that must involve final-state particle $b$ and $c$ quarks. However, for the production of bottom-charmed $\Xi_{bc}$ baryons, the key lies in the nonperturbative hadronization process.

Nonrelativistic quantum chromodynamics (NRQCD) \cite{Bodwin:1994jh} provides a good theoretical framework to effectively solve such production problems \cite{Ma:2003zk}. Within the framework of NRQCD, the production of hadrons can be factorized into two parts: the perturbative region and the nonperturbative region. First, in the hard process, the free final-state particle $b$ and $c$ quarks are perturbatively produced into a diquark with a distinct quantum state $\langle bc\rangle[n]$, and then this intermediate diquark state nonperturbatively transitions to the bottom-charmed baryon $\Xi_{bc}$. This transition process is quantified by long-distance matrix elements (LDMEs) and can be approximated by the wave function at the origin (or its spatial derivative for orbital excited states) within potential models. The spin and color quantum number $[n]$ of $\langle bc\rangle$ diquark can be $[{}^1S_0]_{\bar{\textbf{3}}/\textbf{6}}$, $[{}^3S_1]_{\bar{\textbf{3}}/\textbf{6}}$, $[{}^1P_1]_{\bar{\textbf{3}}/\textbf{6}}$, and $[{}^3P_J]_{\bar{\textbf{3}}/\textbf{6}}$ ($J$=0, 1, 2). These diquark states, respectively, correspond to $^{1} S_{0}$ and $^{3} S_{1}$ configurations of $S$ waves and $^{1} P_{1}$ and $^{3} P_{J}$ configurations for excited $P$ waves. $\overline{\textbf{3}}$ and $\textbf{6}$ are the color quantum numbers of $\langle bc\rangle$ diquarks, representing the color antitriplet state and sextet state for the decomposition of SU(3)$_c$ color group $\mathbf{3}\otimes\mathbf{3}=\bar{\mathbf{3}}\oplus\mathbf{6}$. According to the velocity scaling rule in NRQCD, the orbital angular momentum $L$ yields a suppression $v^{2L}$ in the LDMEs, where $v$ represents the relative velocity between two constituent quarks in the rest frame of the heavy diquark state. So the contribution of $P$ waves ($L$=1) gain a $v^{2}$ suppression against that of $S$ waves ($L$=0). Moreover, the higher angular momentum $D$-wave states carry $v^{4}$ suppression compared to $S$ waves. Hence, the contributions of $D$ waves can be safely neglected in our calculations. In the perturbative hard process, that is, the short-distance coefficient, the energy dependence of the cross section for the photoproduction of $\Xi_{bc}$ in $S$ waves initially increases rapidly, reaches its maximum at around 90 GeV, and then decreases steadily \cite{Ma:2025ito}. Here, in order to give a comprehensive comparison and analysis, the cross section for the photoproduction of the ground and excited state $\Xi_{bc}$ baryons is investigated at ILC and CLIC with collision energy $\sqrt{s}=250, 500, 1000$ GeV, respectively.

In this work, the photoproduction of bottom-charmed baryons is analyzed through the direct channel $\gamma+\gamma \rightarrow \Xi_{bc} +\bar{c}+\bar{b}$ and the resolved channel $\gamma+g \rightarrow \Xi_{bc} +\bar{c}+\bar{b}$ within the NRQCD factorization framework. The rest of the paper is arranged as follows. In
Sec.~\ref{sec:2}, the necessary calculation formulas and technology are provided. The numerical results are presented in Sec.~\ref{sec:3}, involving cross sections, transverse momentum, rapidity, and angular and invariant mass distributions. Section~\ref{sec:4} is reserved for a brief summary.
\section{Calculation Technology}
\label{sec:2}

In the factorization framework of NRQCD, the differential cross section for the photoproduction of $\Xi_{bc}$ can be factorized into
\begin{eqnarray}
	&&\mathrm{d} \sigma(e^{+} e^{-} \rightarrow e^{+} e^{-} +\Xi_{bc}+\bar{c}+\bar{b})
	= \int \mathrm{d} x_{1} f_{\gamma / e}(x_{1}) \int \mathrm{d} x_{2} f_{\gamma / e}(x_{2})\nonumber\\
	&& \times\sum_{i, j} \int \mathrm{d} x_{i} f_{i / \gamma}(x_{i}; \mu_F) \int \mathrm{d} x_{j} f_{j / \gamma}(x_{j}; \mu_F)\nonumber\\
	&& \times  \sum_{n} \mathrm{~d} \hat{\sigma}(i j \rightarrow \langle bc\rangle[n]+\bar{c}+\bar{b}; s_{ij}, \mu_F, \mu_R)\left\langle \mathcal{O}^{\Xi_{bc}}[n]\right\rangle,
\label{faceq}
\end{eqnarray}
where the partonic center-of-mass energy squared is defined as $s_{ij}=x_{i}x_{j}s_{\gamma\gamma}=x_{i}x_{j}x_{1}x_{2}s$, with $\sqrt{s}$ being the $e^{+}e^{-}$ collision energy. $\mu_R$ and $\mu_F$ are the renormalization and factorization scales, respectively.
The energy spectrum of the LBS photon $f_{\gamma/e}(x)$ can be described by Ginzburg et al.~\cite{Ginzburg:1981vm},
\begin{equation}
	f_{\gamma/e}(x)=\frac{1}{N}\left(1-x+\frac{1}{1-x}-\frac{4x}{x_{m}(1-x)} +\frac{4x^2}{x_{m}^2(1-x)^2}\right),
\end{equation}
where $x=E_{\gamma} / E_{e}$ represents the energy fraction of the LBS photon emitted from the initial electron or positron, and the normalization factor $N$ is
\begin{equation}
	N=(1-\frac{4}{x_{m}}-\frac{8}{x_{m}^{2}}) \log(1+x_m)+\frac{1}{2}+\frac{8}{x_{m}}-\frac{1}{2 (1+x_m)^{2}},
\end{equation}
with $x_{m}=4 E_{e} E_{l} \cos ^{2} \frac{\theta}{2}$ ($E_e$ and $E_l$ represent the energies of the incident electron and laser beams, respectively, and the angle between them is denoted as $\theta$). The range of energy fraction $x$ for the LBS photon is constrained by~\cite{Telnov:1989sd}
\begin{equation}
	0 \leq x \leq \frac{x_{m}}{1+x_{m}}(x_m\approx4.83).
\end{equation}

\begin{figure}[!thbp]
\centering
    \includegraphics[scale=0.25]{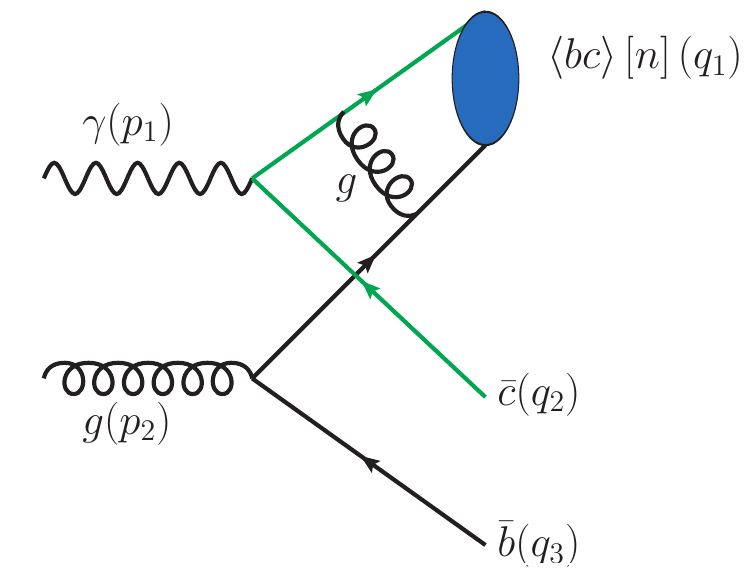}
    \includegraphics[scale=0.25]{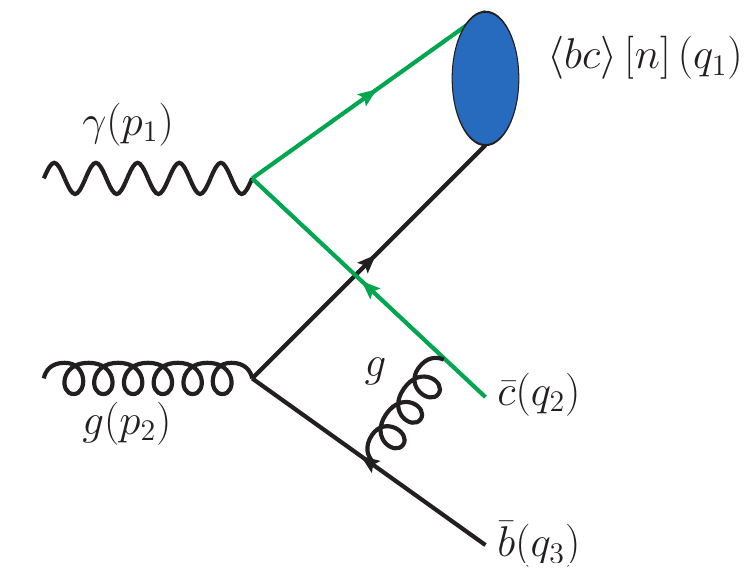}
    \includegraphics[scale=0.25]{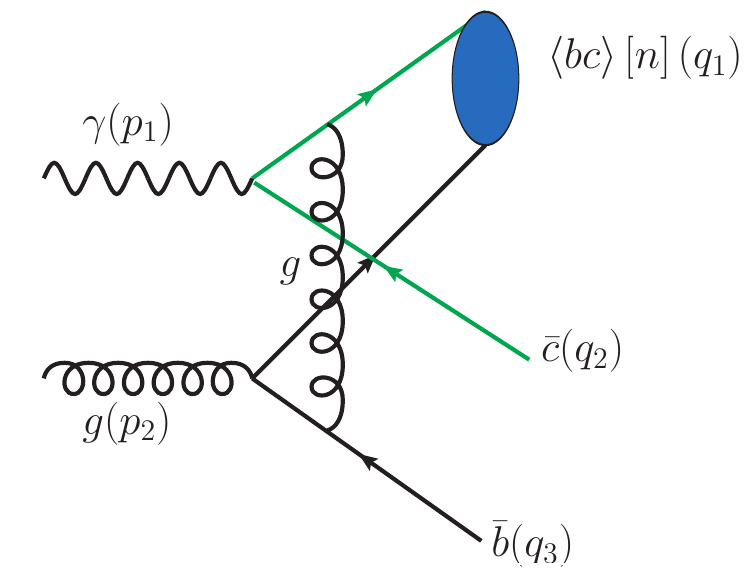}
    \includegraphics[scale=0.25]{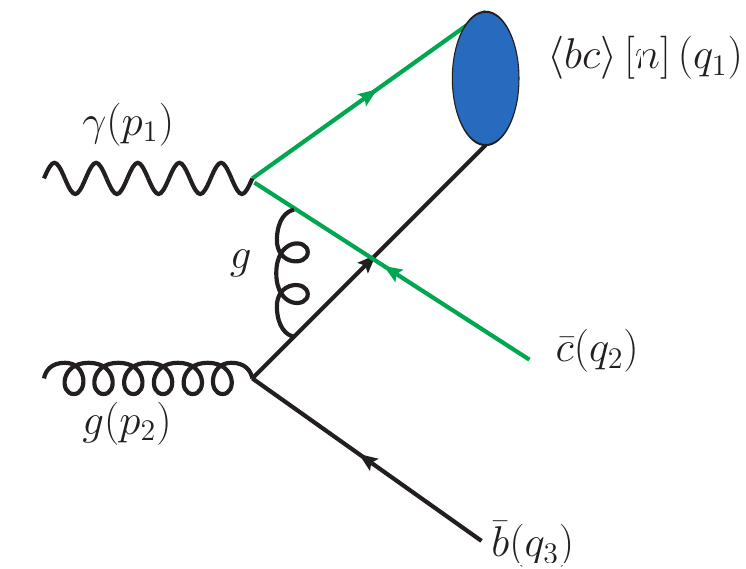}\\
\vspace{0.3cm}    
    \includegraphics[scale=0.25]{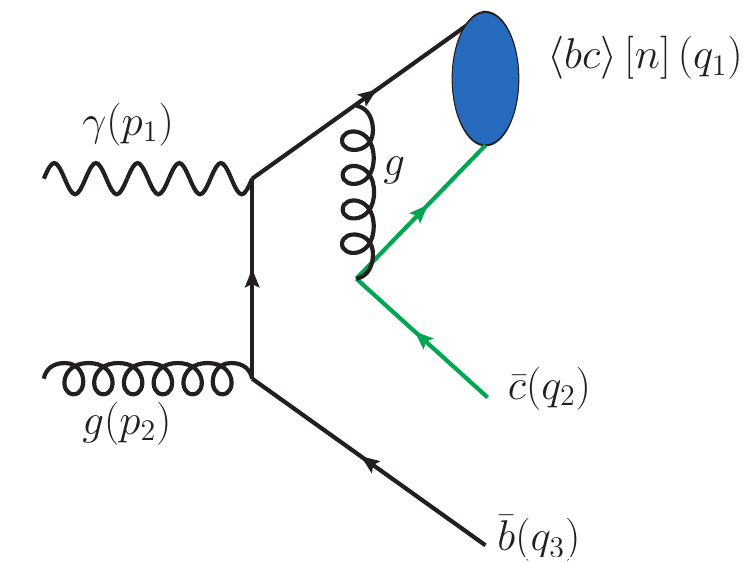}
    \includegraphics[scale=0.25]{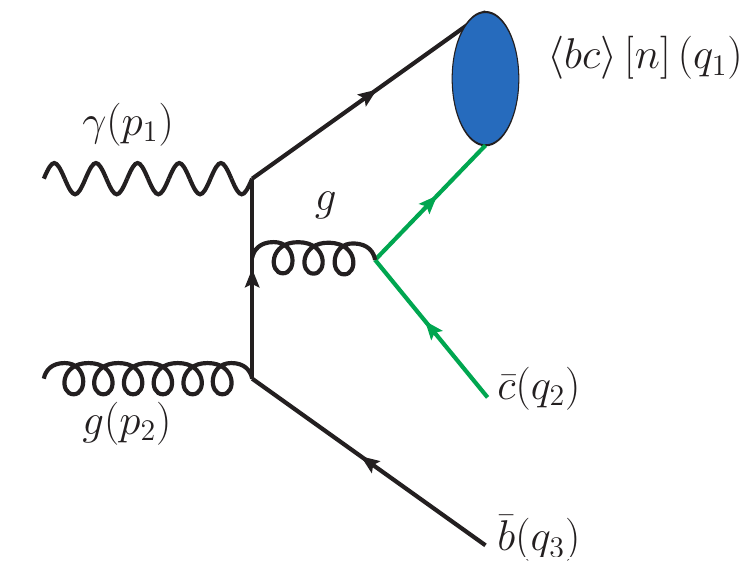}
    \includegraphics[scale=0.25]{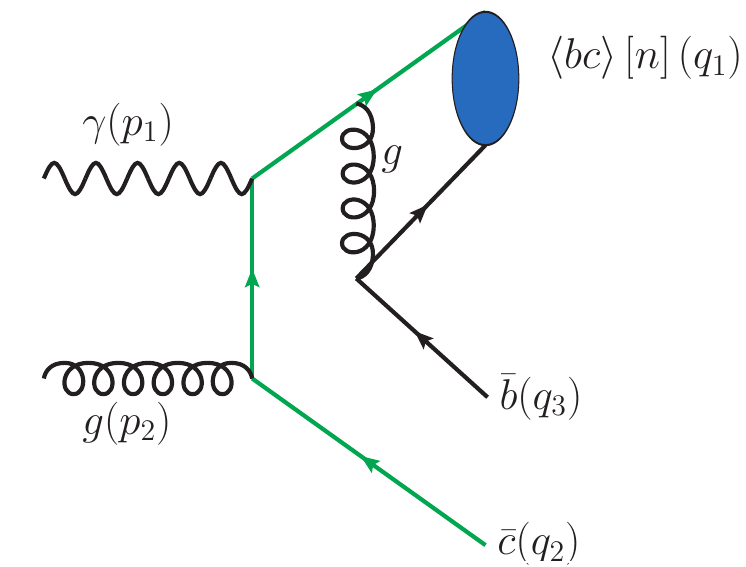}
    \includegraphics[scale=0.25]{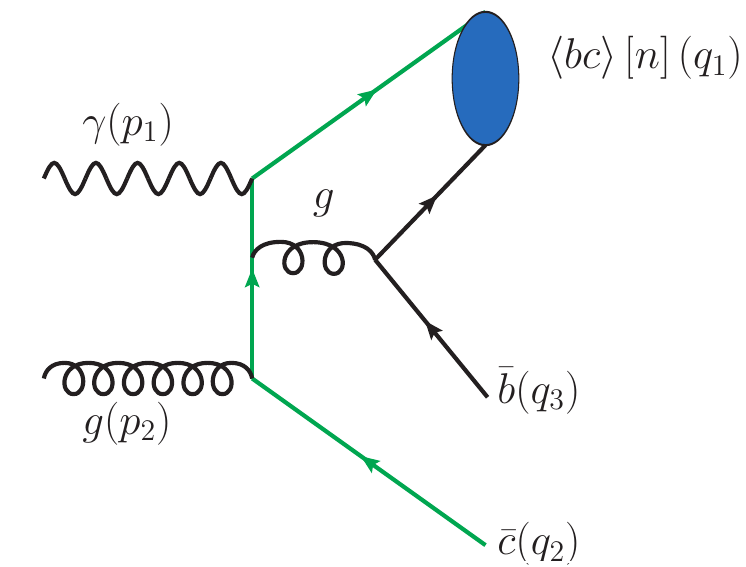}\\
\vspace{0.3cm}        
    \includegraphics[scale=0.22]{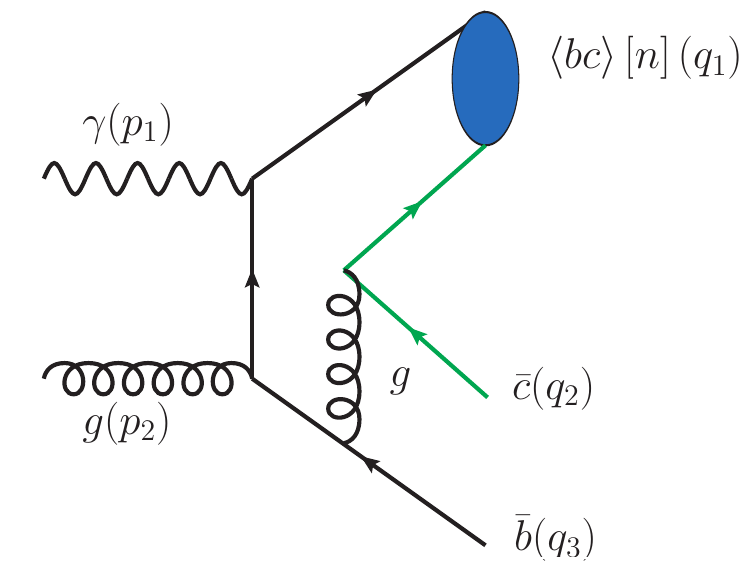}
    \includegraphics[scale=0.22]{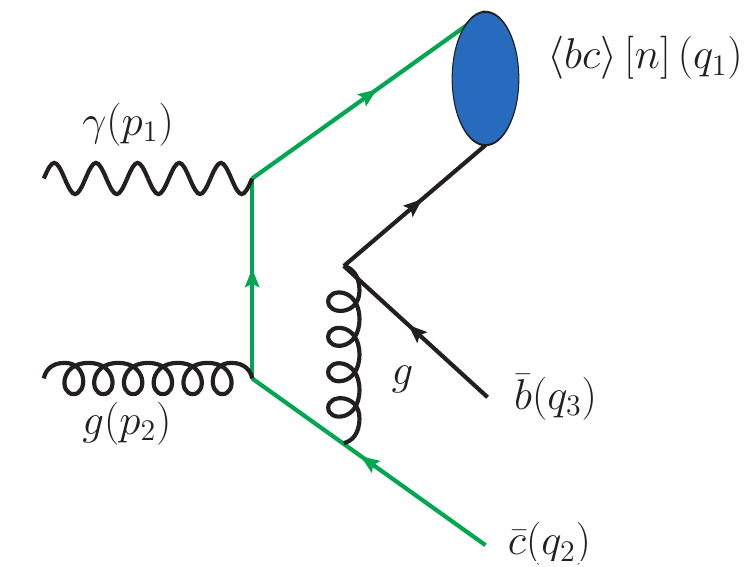}
    \includegraphics[scale=0.25]{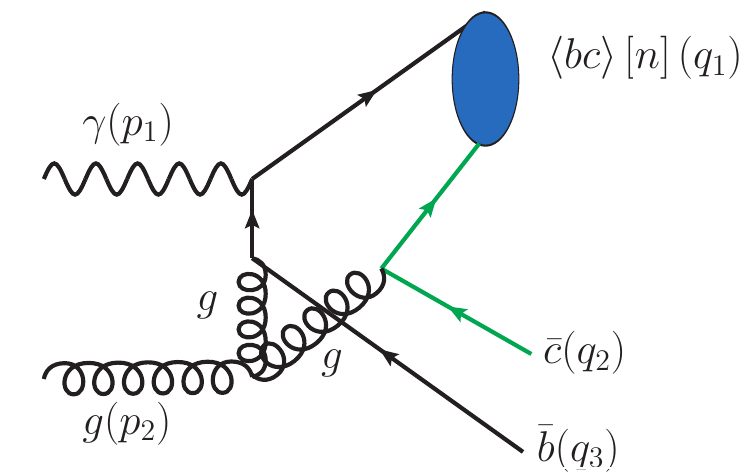}
    \includegraphics[scale=0.25]{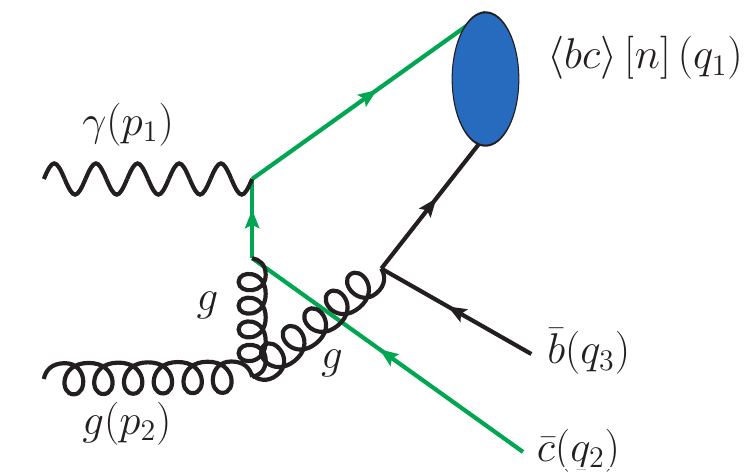}
\caption{Twelve typical Feynman diagrams for the photoproduction of $\Xi_{bc}$ through the intermediate subprocess $\gamma+g\to \langle bc\rangle[n]+\bar{c}+\bar{b}$; another 12 can be obtained by interchanging the initial photon and gluon lines.} 
\label{feynman}
\end{figure}

For two different initial subprocesses, $\gamma+\gamma$ and $\gamma+g$, the Gl\"uck-Reya-Schienbein (GRS) distribution function  $f_{i(j)/\gamma}$ [$i(j)=\gamma,g,u,d,s$] of parton $i(j)$ in the photon~\cite{Gluck:1999ub} is needed, and $f_{\gamma/\gamma}(x)=\delta(1-x)$ is for the direct photoproduction process $\gamma+\gamma$.
The differential partonic cross section $\mathrm{~d} \hat{\sigma}(i j \rightarrow \langle bc\rangle[n]+\bar{c}+\bar{b})$ can be evaluated perturbatively for $\gamma+\gamma$ and $\gamma+g$ subprocesses.
For the partonic processes at leading order, there are 20 Feynman diagrams for $\gamma+\gamma \rightarrow \Xi_{b c}+\bar{c}+\bar{b}$ in ${\cal O}(\alpha^2 \alpha_s^2)$ and 24 diagrams for $\gamma+g \rightarrow \Xi_{bc}+\bar{c}+\bar{b}$ in ${\cal O}(\alpha \alpha_s^3)$.
For $\gamma(p_1)+g(p_2)\to \langle bc\rangle [n](q_1)+\bar{c}(q_2)+\bar{b}(q_3)$, 12 typical Feynman diagrams are shown in Fig. \ref{feynman}, and another 12 can be obtained by interchanging the initial photon and gluon lines. The Feynman diagrams for $\gamma(p_1)+\gamma(p_2)\to \langle bc\rangle [n](q_1)+\bar{c}(q_2)+\bar{b}(q_3)$ can be found in an earlier study \cite{Ma:2025ito}, and they will not be shown here. The additional couplings associated with the partonic content of the resolved photon and with the LBS photon are included in $f_{i/\gamma}$ and $f_{\gamma/e}$, respectively. The GRS parametrization is used for the central predictions. Different photon--parton distribution function (PDF) parametrizations~\cite{Aurenche:2005da, Slominski:2005bw} may introduce an additional normalization and kinematic-shape dependence, especially for the resolved $\gamma+g$ channel through the gluon distribution in the photon.

The differential cross section $\mathrm{~d} \hat{\sigma}(i j \rightarrow \langle bc\rangle[n]+\bar{c}+\bar{b})$ can be rewritten as 
\begin{eqnarray}
\mathrm{~d} \hat{\sigma}(i j \rightarrow \langle bc\rangle[n]+\bar{c}+\bar{b})= \frac{1}{4\sqrt{(p_1\cdot p_2)^2-m^4_{e}}} \overline{\sum}  \big|{\cal M\rm [n]}\big|^{2} \mathrm{d}\Phi_3,
\label{sd-sigma}
\end{eqnarray}
in which $\overline{\Sigma}$ is the average over the spins of the initial states and the sum over colors and spins of the final states, and $\mathrm{d}\Phi_3$ is the three-body phase space. For the production of bottom-charmed baryons $\Xi_{bc}$, the spin and color quantum number $[n]$ of $\langle bc\rangle$-diquark configurations can be
$\langle bc\rangle[^1S_0]_{\bar{\textbf{3}}/\textbf{6}}$, $\langle bc\rangle[^3S_1]_{\bar{\textbf{3}}/\textbf{6}}$, $\langle bc\rangle[^1P_1]_{\bar{\textbf{3}}/\textbf{6}}$, or $\langle bc\rangle[{}^3P_J]_{\bar{\textbf{3}}/\textbf{6}}$. To obtain the amplitude $\cal {M}\rm [n]$, two necessary actions need to be applied. First, the charge conjugation matrix $C=-i \gamma^2 \gamma^0$ needs to be applied to reverse the $c\sim\bar{c}$ fermion chain (green line in Fig. \ref{feynman}). It has been proven that, with the charge conjugation matrix $C$, the amplitude for the production of  $b(q_{11})c(q_{12})+\bar{c}(q_{2})+\bar{b}(q_{3})$ can be related to the amplitude of  $b(q_{11})\bar{c}(q_{12})+c(q_{2})+\bar{b}(q_{3})$ with an additional factor $(-1)^{(n+1)} $ ($n$ is the number of vector vertices in the fermion chain) \cite{Zheng:2015ixa, Zhan:2023jfm, Ma:2025ito}. Here $q_{11}=\frac{m_b}{M_{bc}}q_1+q$ and $q_{12}=\frac{m_c}{M_{bc}}q_1-q$ are the momenta of $b$ and $c$ constituent quarks with $q_1$ being the momentum of the $\langle bc\rangle$ diquark. 
The mass of the $\langle bc\rangle$ diquark is set as $M_{bc} = m_b + m_c$ to ensure that the constituent quarks are strictly on shell, which guarantees the gauge invariance of the amplitude. At the leading order in $v$, the small relative momentum $q$ (with $|\vec{q}|\sim M_{bc}v$) can be set to zero directly in the short-distance amplitude for the production of $S$-wave diquarks.
The second step is to introduce the spin projector $\Pi_{q_1}(q)[n]$ to transform the $b(q_{11})\bar{c}(q_{12})$ constituent quarks into bound states $\langle b\bar{c}\rangle(q_{1})$ and it takes the form of \cite{Petrelli:1997ge}
\begin{equation}
	\Pi_{q_1}(q)[n]=\frac{1}{2\sqrt{M_{bc}}}\varepsilon[n](\slashed {q_{1}}+M_{bc}),
\end{equation}
for $\varepsilon[^1S_0]=\gamma^5$ and $\varepsilon[^3S_1]=\slashed{\varepsilon}$ with the polarization vector $\varepsilon^\beta$ of $^3S_1$ diquark state. The amplitudes of $P$-wave $\Xi_{bc}$ can be obtained by the first derivation of the relative momentum $q$ between two constituent quarks in $S$-wave amplitude, 
\begin{eqnarray}
	\cal{M}\rm[{ }^1 P_1]=\, & \varepsilon_\alpha^l(q_1) \frac{d}{d q_\alpha}\cal M\rm [^1S_0]|_{q=0},\\
	\cal{M}\rm[{ }^3 P_J]=\, & \varepsilon_{\alpha \beta}^J(q_1) \frac{d}{d q_\alpha}\cal M^{\beta}\rm [^3S_1]|_{q=0},
\end{eqnarray}
in which $q$ is always included in the momenta of the constituent quark of the diquark, $q_{11}$ and $q_{12}$, and appears in the spin projector $\Pi_{q_1}$ and propagators. $\varepsilon_\alpha^l(q_1)$ and $\varepsilon_{\alpha \beta}^J(q_1)$ correspond to the polarization vector and polarization tensor of the spin singlet and spin triplet $P$-wave states, where $J$ can be 0, 1, or 2.
The summations over polarization vector and tensor satisfy \cite{Petrelli:1997ge}
\begin{eqnarray}
	&&\sum_{l_z} \varepsilon_\alpha^l \varepsilon_{\alpha^{\prime}}^{l *}=\Pi_{\alpha \alpha^{\prime}},\\
	&&\varepsilon_{\alpha \beta}^0 \varepsilon_{\alpha^{\prime} \beta^{\prime}}^{0 *}=\frac{1}{3} \Pi_{\alpha \beta} \Pi_{\alpha^{\prime} \beta^{\prime}},\\
	&&\sum_{J_z} \varepsilon_{\alpha \beta}^1 \varepsilon_{\alpha^{\prime} \beta^{\prime}}^{1 *}=\frac{1}{2}(\Pi_{\alpha \alpha^{\prime}} \Pi_{\beta \beta^{\prime}}-\Pi_{\alpha \beta^{\prime}} \Pi_{\alpha^{\prime} \beta}),\\
	&&\sum_{J_z} \varepsilon_{\alpha \beta}^2 \varepsilon_{\alpha^{\prime} \beta^{\prime}}^{2 *}=\frac{1}{2}(\Pi_{\alpha \alpha^{\prime}} \Pi_{\beta \beta^{\prime}}+\Pi_{\alpha \beta^{\prime}} \Pi_{\alpha^{\prime} \beta})-\frac{1}{3} \Pi_{\alpha \beta} \Pi_{\alpha^{\prime} \beta^{\prime}} ,
\end{eqnarray}
with the definition $\Pi_{\alpha \beta}=-g_{\alpha \beta}+\frac{q_{1 \alpha}q_{1 \beta}}{M_{b c}^2}$. The color factor $\mathcal{C}_{i j, k}$ for all the Feynman diagrams has been extracted from the amplitude and it satisfies $\mathcal{C}_{i j, k}=\mathcal{N} \times \sum_{a, m, n}(T^a)_{m i}(T^a)_{n j} \times G_{m n k}$, where the normalization constant $\mathcal{N}=\sqrt{1/2}$, $i,j,m,n$ are the color indices of four heavy quarks, $k$ denotes the color index of the diquark, and
$G_{m n k}$ corresponds to either the antisymmetric function $\varepsilon_{m n k}$ for the color antitriplet state or the symmetric function $f_{mnk}$ for the sextet state; they satisfy
\begin{eqnarray}
	\varepsilon_{m n k} \varepsilon_{m^{\prime} n^{\prime} k}=\delta_{m m^{\prime}} \delta_{n n^{\prime}}-\delta_{m n^{\prime}} \delta_{n m^{\prime}},\\
	f_{m n k} f_{m^{\prime} n^{\prime} k}=\delta_{m m^{\prime}} \delta_{n n^{\prime}}+\delta_{m n^{\prime}} \delta_{n m^{\prime}}.
\end{eqnarray}

Through the above perturbative QCD calculations, we can obtain the short-distance coefficients for the production of intermediate $\langle bc\rangle$-diquark states. The next step is the crucial hadronization process of the diquark state transitioning to the $\Xi_{bc}$ baryons through the strong interaction. The LDME $\langle{\cal O}^{\Xi_{bc}}[n]\rangle$ in Eq.~(\ref{faceq}) can be described by $h_{\bar{\textbf{3}}}$ and $h_{\textbf{6}}$ for antitriplet and sextet states, which is nonperturbative and accompanied by significant theoretical uncertainty. Typically, $h_{\bar{\textbf{3}}}$ can be obtained by the potential model and approximately correlated to the Schr$\ddot{o}$dinger wave function at the origin for the $S$-wave diquark and the first-derivative wave function at the origin
for $P$-wave states \cite{Falk:1993gb,Kiselev:1994pu,Bagan:1994dy,Baranov:1995rc,Berezhnoy:1998aa}, which can be naturally connected to the (derivative) radial wave function at the origin, i.e.,
\begin{eqnarray}
	h_{\bar{\textbf{3}}}[S]\simeq|\Psi_{bc}(0)|^2=\frac{1}{4\pi}|R_{bc}(0)|^2 , \\
h_{\bar{\textbf{3}}}[P]\simeq|\Psi^{\prime}_{bc}(0)|^2=\frac{3}{4\pi}|R^{\prime}_{bc}(0)|^2. \label{eq:h3}
\end{eqnarray}
As for $h_{\textbf{6}}$, there is no such clear correlation; however, according to NRQCD theory, the doubly heavy baryon $\Xi_{bc}$ can be expanded into a series of Fock states,
\begin{equation}
|\Xi_{bc}\rangle=c_1(v)|(bc)q\rangle+c_2(v)|(bc)qg\rangle+c_3(v)|(bc)qgg\rangle+\cdots,
\label{Fockststes}
\end{equation}
in which $v$ is the relative velocity of the constituent heavy quarks in the baryon rest frame. In the early literature, there are two points of view for accounting for the contributions of each Fock state. One power counting rule indicates that since a light quark can readily emit gluons, those Fock states in Eq.~(\ref{Fockststes}) are of same importance, specifically, $c_1(v) \sim c_2(v) \sim c_3(v)$ \cite{Ma:2003zk}. Thus, both $h_{\textbf{6}}$ and $h_{\bar{\textbf{3}}}$ are assigned equivalent orders, i.e., $h_{\textbf{6}}=h_{\bar{\textbf{3}}}$. While in ``one-gluon-exchange'' models,  the interaction in the antitriplet state is attractive, which changes to be repulsive for the sextet channel, suggesting suppression at least $v^2$ relative to $h_{\bar{\textbf{3}}}$, or too small to be ignored \cite{Ma:2003zk, Zheng:2015ixa}. Therefore, we adopt $h_{\textbf{6}}=0$ as the conservative baseline in the subsequent calculations and treat other choices explicitly as speculative to estimate the contribution of the color sextet.
\section{Numerical results}
\label{sec:3}

In the numerical calculation, the input parameters are listed as follows \cite{Baranov:1995rc, ParticleDataGroup:2024cfk}:
\begin{eqnarray}
&&|R_{bc}(0)| = 0.722 \, \mathrm{GeV^{3/2}},  ~~~~|R^{\prime}_{bc}(0)| = 0.200 \, \mathrm{GeV^{5/2}},  \nonumber\\
&&m_b = 5.1 \, \mathrm{GeV}, ~~~~m_c = 1.8 \, \mathrm{GeV}, ~~~~M_{bc}=m_b+m_c, \nonumber\\
&& \alpha=1/132.23, ~~~~~\mu_R=\mu_F\equiv\mu_0=\sqrt{M^2_{\Xi_{bc}} + p^2_T},
\end{eqnarray}
where $|R_{bc}(0)| $ and $|R^{\prime}_{bc}(0)|$ are evaluated under the $\rm K^2O$ potential motivated by QCD with a three-loop function \cite{Kiselev:2002iy}. For the central prediction, the renormalization and factorization scales are identified and chosen as the transverse mass of $\Xi_{bc}$, $\mu_R=\mu_F=\mu_0=\sqrt{M_{\Xi_{bc}}^2+p_T^2}$. The one-loop running coupling $\alpha_s(\mu_R)$ and the GRS photon PDF evaluated at $\mu_F$ are then used consistently.

At ILC and CLIC with the collision energy $\sqrt{s}=250,~500,~1000 \mathrm{GeV}$, the cross sections of all considered antitriplet intermediate diquark states are conducted via both photon-photon fusion and photon-gluon fusion, and the results are provided in Tables~\ref{csrr} and \ref{csrg}. By summing up all the considered antitriplet states, we obtained the combined contributions of $S$ and $P$ waves, and the results are listed in Table \ref{cstot}. From these three tables, we can see that

\begin{itemize}
 \item With the increasing of $\sqrt s$, the contribution of each $\langle bc\rangle$-diquark state decreases via $\gamma$+$\gamma$ fusion. However, for $\gamma$+$g$ fusion, the contribution of each $\langle bc\rangle$-diquark state increases as the energy increases. Based on the comprehensive analysis of the two subprocesses of photoproduction, the total cross section increases with the increase of collision energy. At $\sqrt{s}=250\,\mathrm{GeV}$, the contribution of $\gamma+\gamma$ fusion is dominant in the photoproduction process. When the collision energy reaches 500 GeV or higher, the contribution of photon-gluon fusion shows a clearly overwhelming trend.
 \item Among these diquark states, the contribution of $[^3S_1]_{\bar{\textbf{3}}}$ is the largest, accounting for  63.50\%, 63.67\%, and 63.95\% (71.62\%, 71.20\%, and 71.04\%) of the total in $\gamma$+$\gamma$ ($\gamma$+$g$) fusion when $\sqrt s=$250, 500, 1000 GeV. The $[^3P_2]_{\bar{\textbf{3}}}$ state dominates the total $P$-wave state, accounting for 50.47\% (50.46\% and 51.06\%) and 50.48\% (50.92\% and 50.77\%) in $\gamma$+$\gamma$ and $\gamma$+$g$ fusion when $\sqrt s=$250 (500 and 1000) GeV, respectively.
 \item The contribution of $S$-wave diquark states is significantly larger than that of the $P$ wave. For the subprocess $\gamma$+$\gamma$ ($\gamma$+$g$) fusion when $\sqrt s=$250, 500, and 1000 GeV, the contribution of the $P$-wave state is 8.44\%, 8.81\%, and 8.87\% (7.41\%, 7.46\%, and 7.43\%) of that of the $S$ wave.
 \item The combined $(S\text{-wave},P\text{-wave},\mathrm{total})$ cross sections are
(39.54, 3.19, 42.73), (41.61, 3.27, 44.88), and (57.82, 4.37, 62.19)~$\mathrm{fb}$ for $\sqrt{s}=250$, 500, and 1000~$\mathrm{GeV}$, respectively. The corresponding ratios $\sigma_{P-\rm wave}/\sigma_{S-\rm wave}$ are 8.07\%, 7.86\%, and 7.56\%, while $\sigma_{P-\rm wave}/\sigma_{\rm tot}$ are 7.47\%, 7.29\%, and 7.03\%.
 \end{itemize}

Based on the analysis of two subprocesses of photoproduction, the contributions of the $P$-wave state is approximately 8\% of the $S$-wave production. Assuming that at ILC and CLIC, with an illustrative integrated luminosity of $\mathcal{O}(10^4)\mathrm{~fb^{-1}}$~\cite{Chen:2014frw,Zhan:2023vwp,Zhan:2023jfm}, the produced events of $P$-wave excited baryons $\Xi_{bc}$ from two subprocesses of photoproduction can be approximately $3.19\times10^4$ ($3.27\times10^4$, $4.37\times10^4$) at $\sqrt{s}=250~(500, 1000)\mathrm{~GeV}$.
Subsequently, if the $P$-wave states can decay to the ground state with a 100\% probability, approximately $4.27\times10^5$ ($4.49\times10^5$, $6.22\times10^5$) of baryons could be produced at $\sqrt{s}=250~(500, 1000)\mathrm{~GeV}$, providing an upper-bound estimation for the feed-down contributions. Note that these estimated events are just parton-level inclusive rates. To evaluate the actual experimental discovery potential, one must further consider the realistic detector acceptance, hadronization efficiencies, and physical background subtractions.

\begin{table}[htp]
\caption{The cross sections $\sigma$ (in units of femtobarn) of all considered antitriplet intermediate diquark states for $\Xi_{bc}$ photoproduction via photon-photon fusion using LBS spectrum at ILC and CLIC with different collision energy $\sqrt{s}$~$(\mathrm{GeV})$.}
\centering
\begin{tabular}{|c|c|c|c|c|c|c|}
\hline
\diagbox{$\sqrt {s}$}{$\sigma(n)$}  & $[^1S_0]_{\bar{\textbf{3}}}$ &  $[^3S_1]_{\bar{\textbf{3}}}$ &  $[^1P_1]_{\bar{\textbf{3}}}$  & $[^3P_0]_{\bar{\textbf{3}}}$ & $[^3P_1]_{\bar{\textbf{3}}}$  & $[^3P_2]_{\bar{\textbf{3}}}$ \\
\hline \hline
250 &   7.90     &   17.47   &   0.40          &   0.33      &   0.33     &  1.08   \\
\hline
500 &  3.80     &   8.57     &    0.21        &    0.17      &   0.16   &  0.55   \\
\hline
1000 &   1.61   &   3.69    &    0.09          &   0.07     &   0.07   &  0.24  \\
\hline                   
\end{tabular}
\label{csrr} 
\end{table}

\begin{table}[htp]
\caption{The cross sections $\sigma$ (in units of femtobarn) of all considered antitriplet intermediate diquark states for $\Xi_{bc}$ photoproduction via photon-gluon fusion using LBS spectrum at ILC and CLIC with different collision energy $\sqrt{s}$~$(\mathrm{GeV})$.}
\centering
\begin{tabular}{|c|c|c|c|c|c|c|c|}
\hline
\diagbox{$\sqrt {s}$}{$\sigma(n)$}  & $[^1S_0]_{\bar{\textbf{3}}}$ & $[^3S_1]_{\bar{\textbf{3}}}$ & $[^1P_1]_{\bar{\textbf{3}}}$ & $[^3P_0]_{\bar{\textbf{3}}}$ & $[^3P_1]_{\bar{\textbf{3}}}$  & $[^3P_2]_{\bar{\textbf{3}}}$ \\ 
\hline \hline
250 &   3.27    &   10.90   &    0.28           &   0.08     &   0.16     &  0.53    \\
\hline
500 &   6.87    &   22.37     &    0.57         &   0.15     &   0.35     &  1.11    \\ 
\hline
1000 &   12.44   &   40.08    &    1.03       &   0.27     &   0.62    &  1.98     \\
\hline                     
\end{tabular}
\label{csrg} 
\end{table}

\begin{table}
\caption{The total cross sections $\sigma$ (in units of femtobarn) for $\Xi_{bc}$ photoproduction via photon-photon and photon-gluon fusion using LBS spectrum at ILC and CLIC with different collision energy $\sqrt{s}$~$(\mathrm{GeV})$.}
\centering
\begin{tblr}{
  cells = {c},
  cell{1}{1} = {r=2}{},
  cell{1}{2} = {c=3}{},
  cell{1}{5} = {c=3}{},
  cell{1}{8} = {r=2}{},
  vlines,
  hline{1,3-6} = {-}{},
  hline{2} = {2-7}{},
}
$\sqrt s$ & $\gamma+\gamma$   &       &       & $\gamma+g$   &       &       & Total \\
        & $S$ wave & $P$ wave & Total & $S$ wave & $P$ wave & Total &       \\
250     &   25.37    &  2.14    &  27.51   &  14.17  & 1.05  &  15.22  & 42.73    \\
500     &  12.37  &  1.09  &  13.46  &  29.24  & 2.18  &  31.42 & 44.88 \\
1000    &  5.30  &  0.47  & 5.77  &  52.52  & 3.90 &  56.42 &  62.19
\end{tblr}
\label{cstot} 
\end{table}

\begin{figure}
	\centering 
	\includegraphics[width=.32\textwidth]{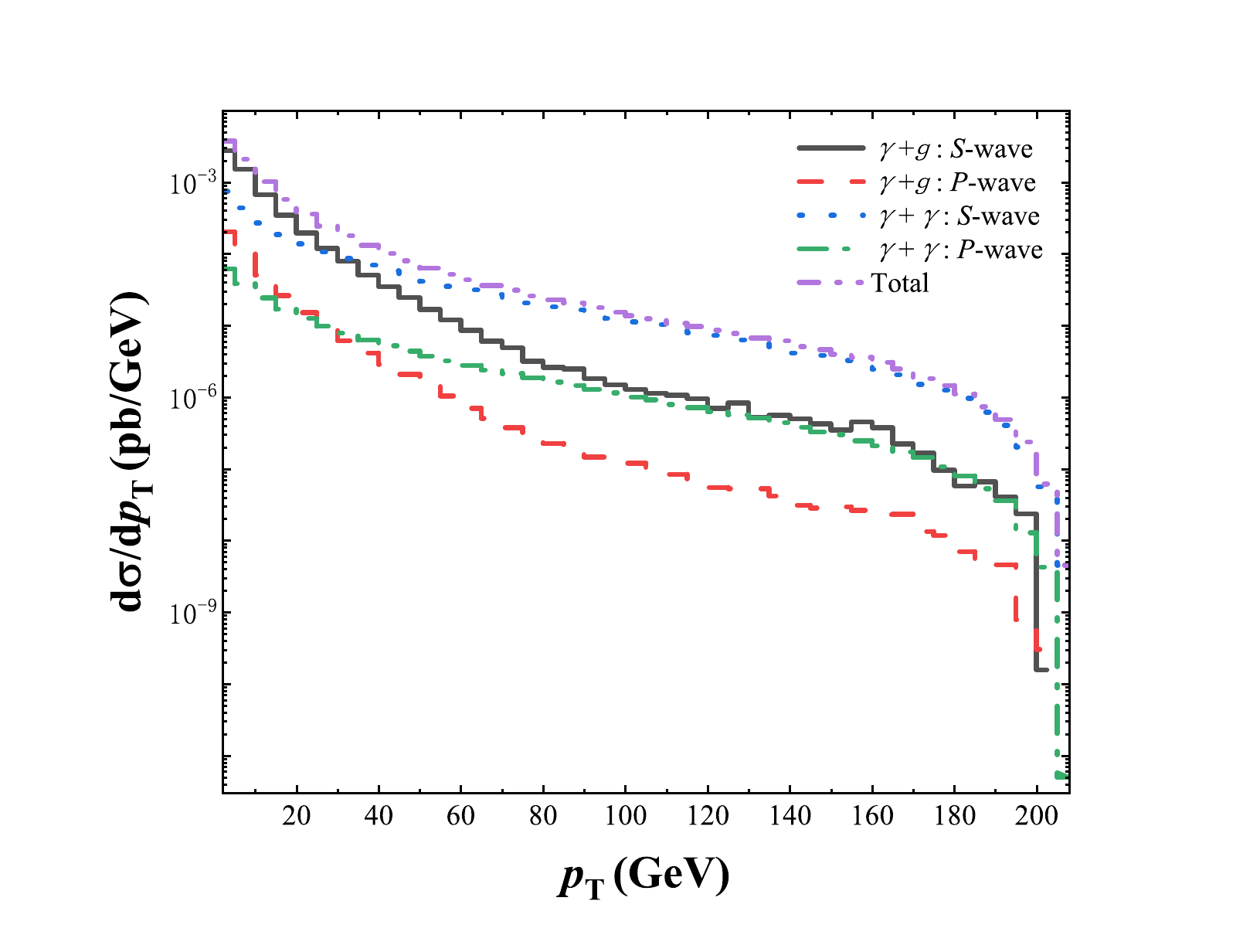}
	\includegraphics[width=.32\textwidth]{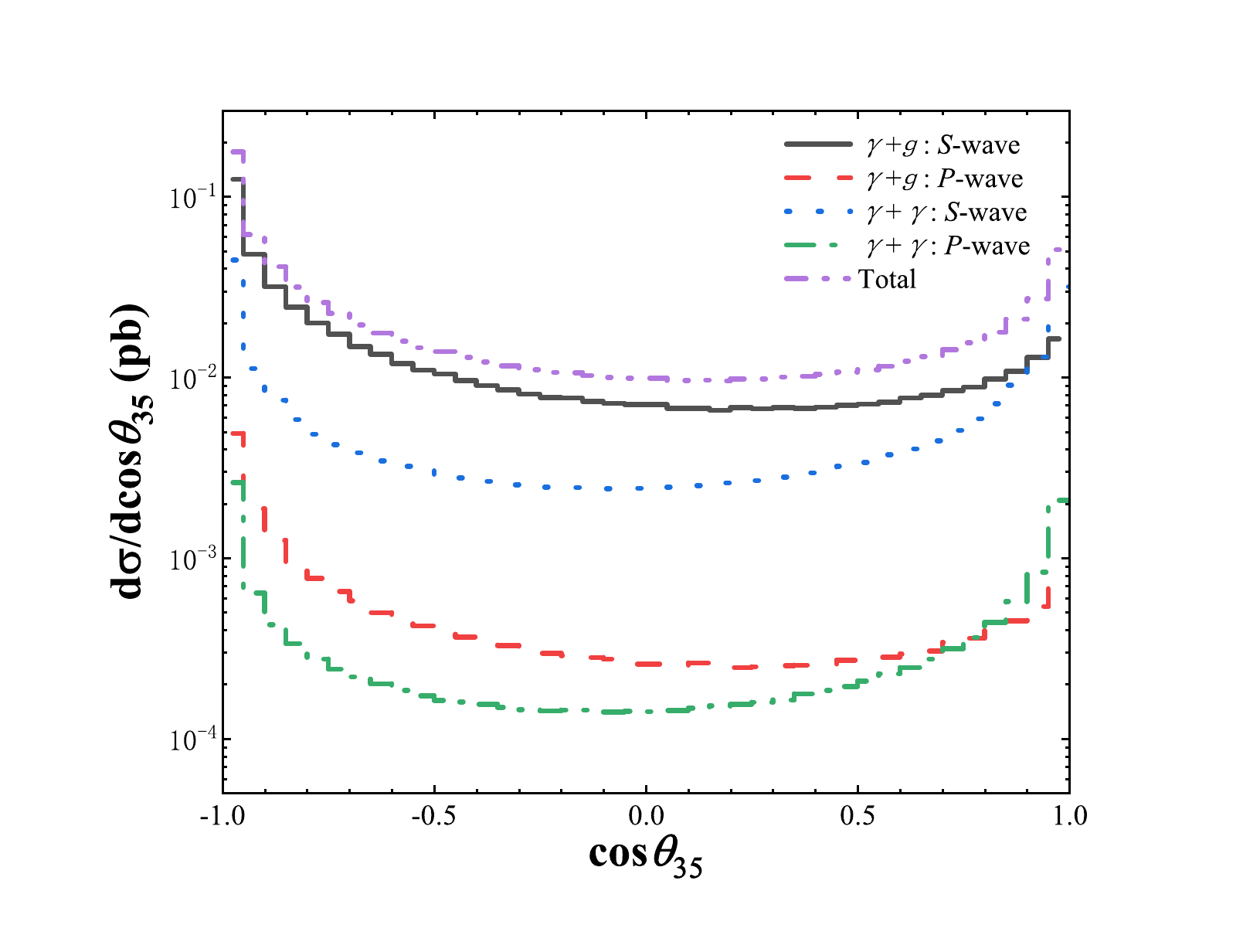}	
	\includegraphics[width=.32\textwidth]{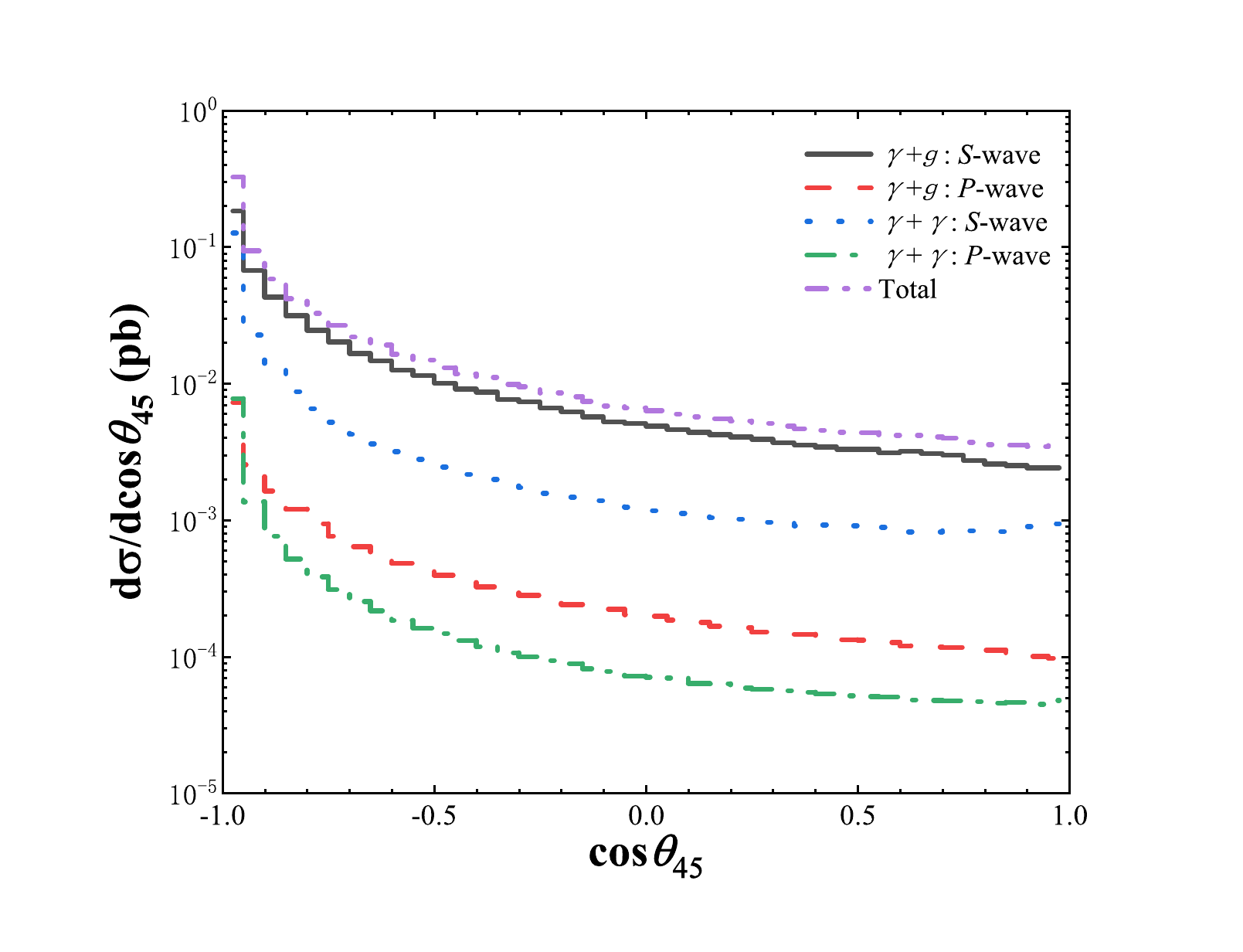}\\
	\includegraphics[width=.31\textwidth]{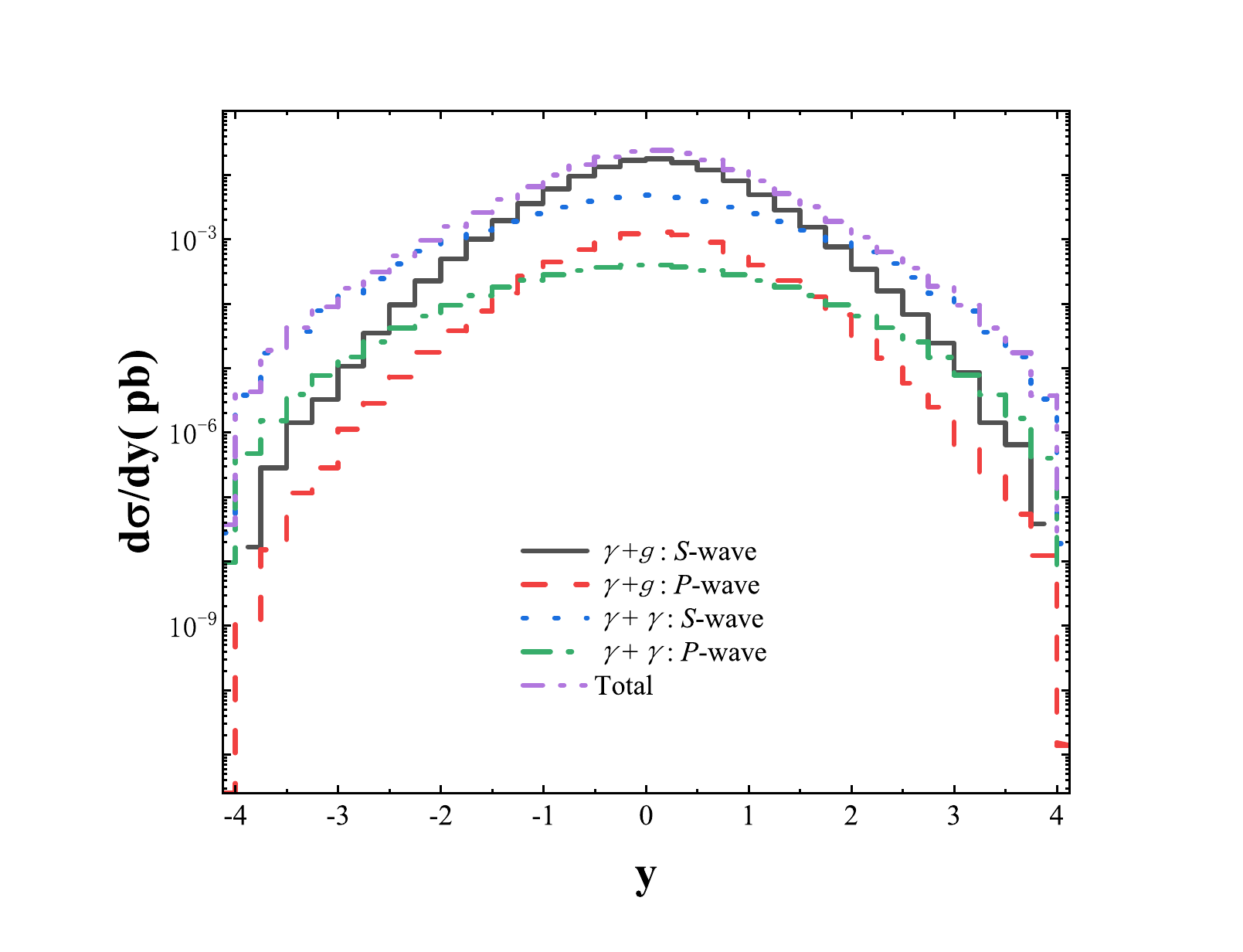}
	\includegraphics[width=.31\textwidth]{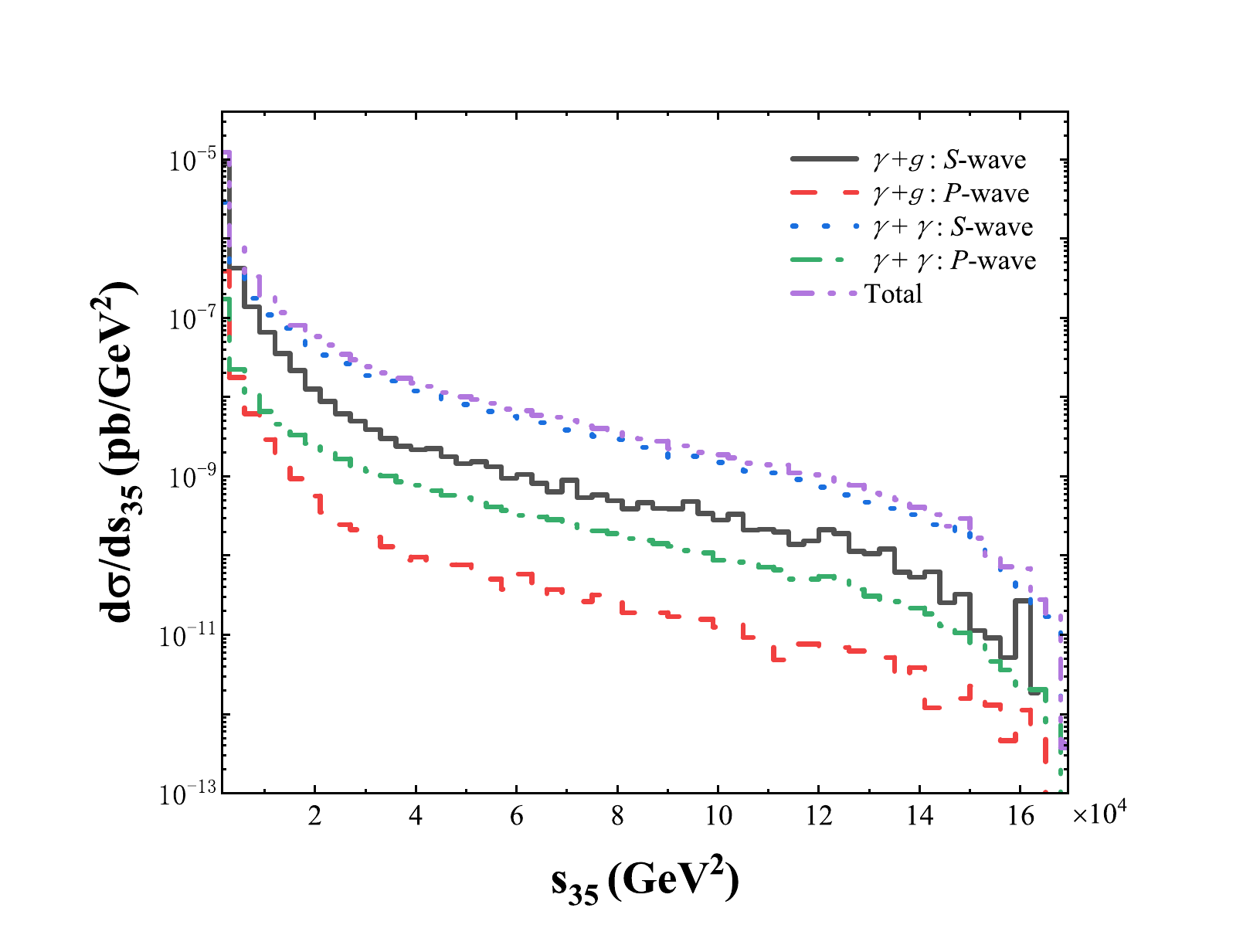}	
	\includegraphics[width=.31\textwidth]{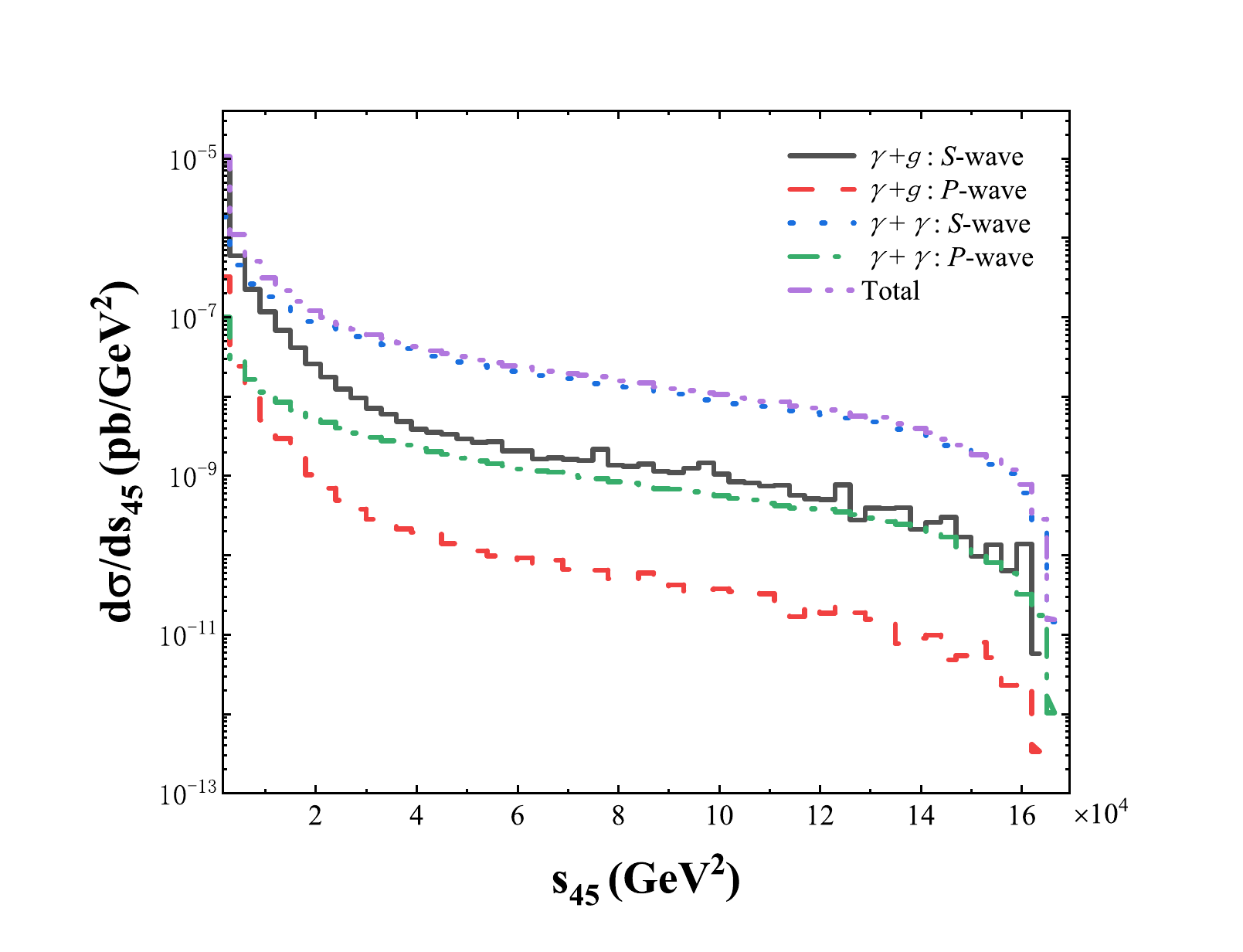}	
	\caption{\label{distribution} Kinematic distributions for the photoproduction of $\Xi_{bc}$ at ILC and CLIC with $\sqrt{s}=$$500$ $\mathrm{GeV}$.}
\end{figure}

To reveal more kinematic characteristics for the photoproduction of $\Xi_{bc}$ at ILC and CLIC ($\sqrt{s}=500\,\mathrm{GeV}$ is selected as a representative), the differential distributions are presented in Fig.\ref{distribution}, including the transverse momentum ($p_T$) and the rapidity ($y$) of $\Xi_{bc}$ in diquark multistates, the angular ($\cos\theta_{35}$ and $\cos\theta_{45}$) and the invariant mass ($s_{35}$ and $s_{45}$) distributions. Here the definitions of the invariant mass are $s_{35}= (q_1 + q_3)^2$ and $s_{45}= (q_2 + q_3)^2$, and $\theta_{35}$ ($\theta_{45}$)  is the angle between the momenta $q_1$ ($q_2$) and $q_3$. Because of the excessive intermediate $\langle bc\rangle$-diquark states in the two subprocesses of photoproduction and the similar trends of each diquark state in $S$ and $P$ waves, the contribution of each diquark state in $S$ and $P$ waves is summed when presenting the differential distributions. To analyze the overall distribution, the contributions of $S$ and $P$ waves are summed up and labeled as ``Total.'' Six different distributions all show that the contribution of the $S$ wave is clearly dominant throughout the whole region compared with the $P$ wave of each subprocess. The distribution of transverse momentum $p_T$ in Fig.\ref{distribution} reveals that, both in $S$ and $P$ waves, the photoproduction of $\Xi_{bc}$ through $\gamma+g$ fusion is larger than that via the photon-photon fusion channel in the region of small $p_T$. Such behavior is mainly due to the rapid growth of the gluon distribution $f_{g/\gamma}(x)$ in the small-$x$ region, combined with the collinear enhancement from the triple-gluon vertex diagrams at small $p_T$.
However, when $p_T$ is greater than about 25 GeV, a significant reverse increase occurs. The overall trend of total $S$ and $P$ waves shows a downward trend, and there are significantly more events in the small-$p_T$ region. For the rapidity distribution for the photoproduction, two subprocesses are concentrated within the range of $[-4.2,+4.2]$, and the curve of $\gamma+g$ has a steeper slope, while that of $\gamma+\gamma$ is more gradual. It can be seen from the angular distributions $\cos\theta_{45}$ that when the angle is $\pi$, that is, when the two heavy antiquarks move back to back, the contribution is the greatest. $\cos\theta_{35}$ indicates that the contribution is significant when the angle is 0 or $\pi$, meaning that the bottom-charmed baryon is most likely to move parallel to the two heavy antiquarks. The same conclusion has also been evident in the distributions of the invariant masses $s_{35}$ and $s_{45}$.

Subsequently, the heavy-quark-mass dependence and the renormalization-scale dependence of the cross sections are quantified. The color sextet LDME $h_{\mathbf{6}}$ is treated separately as a hadronization model benchmark rather than as a statistically defined perturbative uncertainty. In addition, the choice of photon PDF constitutes an unquantified systematic dependence of the resolved channel, as noted in Sec.~\ref{sec:2}. As constituent quarks, the masses $m_b$ and $m_c$ of heavy quarks play a crucial role for the search of doubly heavy baryon $\Xi_{bc}$. The choice of heavy quark masses introduces significant theoretical uncertainty. We analyze the theoretical uncertainty of heavy quark masses by varying $m_c = 1.8 \pm 0.3$ and $m_b = 5.1 \pm 0.3$ GeV, and the results are shown in Tables \ref{tab:mass_uncertainty_gammagamma} and \ref{tab:mass_uncertainty_color3bar}, corresponding to the direct and resolved photoproduction of $\Xi_{bc}$ baryons. When varying $m_c$ ($m_b$), the other heavy quark mass is fixed at $m_b=5.1~\mathrm{GeV}$ ($m_c=1.8~\mathrm{GeV}$). In Tables \ref{tab:mass_uncertainty_gammagamma} and \ref{tab:mass_uncertainty_color3bar}, $\sigma_S$ is the contribution of $S$-wave states, $[{}^1S_0]_{\bar{\mathbf 3}}$ and $[{}^3S_1]_{\bar{\mathbf 3}}$, and $\sigma_P$ is the contribution of $P$-wave states, $[{}^1P_1]_{\bar{\mathbf 3}}$ and $[{}^3P_J]_{\bar{\mathbf 3}}$ $(J=0,1,2)$. $\sigma_{tot}$ is the total contribution of $S$- and $P$-wave states for the photoproduction of  $\Xi_{bc}$. One can find that the cross section of each intermediate diquark state has a strong dependence on the heavy quark masses $m_c$ and $m_b$, and it shows a monotonically decreasing trend as the heavy quark mass increases, mainly due to the inhibition of phase space. 
Regarding the photoproduction channel $\gamma+\gamma$ ($\gamma+g$), the total cross section $\sigma_{tot}$ can vary by $-30.24\% \sim+54.68\%$ 
($-36.86\% \sim+73.23\%$) for $m_c = 1.8 \pm 0.3$ GeV and  $-11.74\% \sim +13.74\%$  ($-17.06\% \sim +21.96\%$)  for $m_b = 5.1 \pm 0.3$ GeV. The contribution of $P$-wave states can be $6.83\%\sim12.00\%$  ($5.70\%\sim 10.09\%$)  of that of the $S$-wave states and $6.39\%\sim10.71\%$  ($5.39\%\sim9.17\%$) of the total cross section with the uncertainty of heavy quark mass.

\begin{table}[tbp]
\caption{Heavy-quark-mass dependence of the color antitriplet contribution for the direct photoproduction channel $\gamma+\gamma\to \Xi_{bc}+\bar c+\bar b$ at $\sqrt{s}=500~\mathrm{GeV}$. The cross sections are given in units of femtobarn.}
\label{tab:mass_uncertainty_gammagamma}
\begin{ruledtabular}
\begin{tabular}{lccccc}
State
& $m_c=1.5$
& $(m_c,m_b)=(1.8,5.1)$
& $m_c=2.1$
& $m_b=4.8$
& $m_b=5.4$ \\
\hline
$[{}^1S_0]_{\bar{\mathbf 3}}$
& 5.62 & 3.80 & 2.74 & 4.32 & 3.35 \\
$[{}^3S_1]_{\bar{\mathbf 3}}$
& 12.97 & 8.57 & 6.05 & 9.73 & 7.59 \\
$[{}^1P_1]_{\bar{\mathbf 3}}$
& 0.44 & 0.21 & 0.11 & 0.24 & 0.18 \\
$[{}^3P_0]_{\bar{\mathbf 3}}$
& 0.31 & 0.17 & 0.10 & 0.20 & 0.14 \\
$[{}^3P_1]_{\bar{\mathbf 3}}$
& 0.33 & 0.16 & 0.09 & 0.19 & 0.14 \\
$[{}^3P_2]_{\bar{\mathbf 3}}$
& 1.15 & 0.55 & 0.30 & 0.63 & 0.48 \\
\hline
$\sigma_S$
& 18.59 & 12.37 & 8.79 & 14.05 & 10.94 \\
$\sigma_P$
& 2.23 & 1.09 & 0.60 & 1.26 & 0.94 \\
$\sigma_{\rm tot}$
& 20.82 & 13.46 & 9.39 & 15.31 & 11.88 \\
$\sigma_P/\sigma_S~(\%)$
& 12.00 & 8.81 & 6.83 & 8.97 & 8.59 \\
$\sigma_P/\sigma_{\rm tot}~(\%)$
& 10.71 & 8.10 & 6.39 & 8.23 & 7.91 \\
\end{tabular}
\end{ruledtabular}
\end{table}

\begin{table}[tbp]
\caption{Heavy-quark-mass dependence of the color antitriplet contribution for the resolved photoproduction channel $\gamma+g\to \Xi_{bc}+\bar c+\bar b$ at $\sqrt{s}=500~\mathrm{GeV}$. The cross sections are given in units of femtobarn.}
\label{tab:mass_uncertainty_color3bar}
\begin{ruledtabular}
\begin{tabular}{lccccc}
State & $m_c=1.5$ & $(m_c,m_b)=(1.8,5.1)$ & $m_c=2.1$ & $m_b=4.8$ & $m_b=5.4$ \\
\hline
$[{}^1S_0]_{\bar{\mathbf 3}}$ & 11.75 & 6.87 & 4.36 & 8.35 & 5.72 \\
$[{}^3S_1]_{\bar{\mathbf 3}}$ & 37.69 & 22.37 & 14.41 & 27.28 & 18.59 \\
$[{}^1P_1]_{\bar{\mathbf 3}}$ & 1.32 & 0.57 & 0.28 & 0.71 & 0.46 \\
$[{}^3P_0]_{\bar{\mathbf 3}}$ & 0.32 & 0.15 & 0.08 & 0.19 & 0.12 \\
$[{}^3P_1]_{\bar{\mathbf 3}}$ & 0.80 & 0.35 & 0.17 & 0.42 & 0.28 \\
$[{}^3P_2]_{\bar{\mathbf 3}}$ & 2.55 & 1.11 & 0.54 & 1.37 & 0.89 \\
\hline
$\sigma_S$ & 49.44 & 29.24 & 18.77 & 35.63 & 24.31 \\
$\sigma_P$ & 4.99 & 2.18 & 1.07 & 2.69 & 1.75 \\
$\sigma_{\rm tot}$ & 54.43	 & 31.42 & 19.84 & 38.32 & 26.06 \\
$\sigma_P/\sigma_S~(\%)$ & 10.09 & 7.46 & 5.70 & 7.55 & 7.20 \\
$\sigma_P/\sigma_{\rm tot}~(\%)$ & 9.17 & 6.94 & 5.39 & 7.02 & 6.72 \\
\end{tabular}
\end{ruledtabular}
\end{table}

For the partonic processes at leading order, the direct photoproduction channel $\gamma+\gamma \rightarrow \Xi_{b c}+\bar{c}+\bar{b}$ is in ${\cal O}(\alpha^2 \alpha_s^2)$ and the resolved photoproduction channel $\gamma+g \rightarrow \Xi_{bc}+\bar{c}+\bar{b}$ is in ${\cal O}(\alpha \alpha_s^3)$. To assess the residual scale dependence, the common scale $\mu_R=\mu_F\equiv\mu$ is varied from $\mu_0/2$ to $2\mu_0$ with $\mu_0=\sqrt{M_{\Xi_{bc}}^2+p_T^2}$ at $\sqrt{s}=500~\mathrm{GeV}$. The corresponding results for the direct and resolved channels are presented in Tables~\ref{tab:scale_uncertainty_gammagamma} and \ref{tab:scale_uncertainty_color3bar}, respectively. The cross sections are given in units of femtobarn. Clearly, due to the running behavior of the strong coupling constant with the energy scale, each state and the overall result are decreasing as the renormalization scale increases; the theoretical uncertainty of the total contribution $\sigma_{tot}$ is approximately $-26.67\%$ to $+46.73\%$ ( $-13.30\%$ to $+16.01\%$) for $\gamma+\gamma$ ($\gamma+g$) with $\mu \in [\mu_0/2, 2\mu_0]$. However, the contribution of the $P$-wave states relative to that of the $S$-wave states and the total cross section has a relatively stable proportion.

\begin{table}[tbp]
\caption{Common renormalization-/factorization-scale dependence of the color antitriplet contribution for the direct photoproduction channel $\gamma+\gamma\to \Xi_{bc}+\bar c+\bar b$ at $\sqrt{s}=500~\mathrm{GeV}$ with $(m_c,m_b)=(1.8,5.1)~\mathrm{GeV}$.}
\label{tab:scale_uncertainty_gammagamma}
\begin{ruledtabular}
\begin{tabular}{lccc}
State
& $\mu=\mu_0/2$
& $\mu=\mu_0$
& $\mu=2\mu_0$ \\
\hline
$[{}^1S_0]_{\bar{\mathbf 3}}$  & 5.57 & 3.80 & 2.79 \\
$[{}^3S_1]_{\bar{\mathbf 3}}$  & 12.58 & 8.57 & 6.29 \\
$[{}^1P_1]_{\bar{\mathbf 3}}$  & 0.30 & 0.21 & 0.15 \\
$[{}^3P_0]_{\bar{\mathbf 3}}$  & 0.25 & 0.17 & 0.12 \\
$[{}^3P_1]_{\bar{\mathbf 3}}$  & 0.24 & 0.16 & 0.12 \\
$[{}^3P_2]_{\bar{\mathbf 3}}$  & 0.81 & 0.55 & 0.40 \\
\hline
$\sigma_S$
& 18.15 & 12.37 & 9.08 \\
$\sigma_P$
& 1.60 & 1.09 & 0.79 \\
$\sigma_{\rm tot}$
& 19.75 & 13.46 & 9.87 \\
$\sigma_P/\sigma_S~(\%)$
& 8.82 & 8.81 & 8.70 \\
$\sigma_P/\sigma_{\rm tot}~(\%)$
& 8.10 & 8.10 & 8.00 \\
\end{tabular}
\end{ruledtabular}
\end{table}

\begin{table}[tbp]
\caption{Common renormalization-/factorization-scale dependence of the color antitriplet contribution for the resolved photoproduction channel $\gamma+g\to \Xi_{bc}+\bar c+\bar b$ at $\sqrt{s}=500~\mathrm{GeV}$ with $(m_c,m_b)=(1.8,5.1)~\mathrm{GeV}$.}
\label{tab:scale_uncertainty_color3bar}
\begin{ruledtabular}
\begin{tabular}{lccc}
State & $\mu=\mu_0/2$ & $\mu=\mu_0$ & $\mu=2\mu_0$ \\
\hline
$[{}^1S_0]_{\bar{\mathbf 3}}$ & 7.99 & 6.87 & 5.94 \\
$[{}^3S_1]_{\bar{\mathbf 3}}$ & 25.94 & 22.37 & 19.42 \\
$[{}^1P_1]_{\bar{\mathbf 3}}$ & 0.66 & 0.57 & 0.50 \\
$[{}^3P_0]_{\bar{\mathbf 3}}$ & 0.18 & 0.15 & 0.13 \\
$[{}^3P_1]_{\bar{\mathbf 3}}$ & 0.40 & 0.35 & 0.30 \\
$[{}^3P_2]_{\bar{\mathbf 3}}$ & 1.28 & 1.11 & 0.95 \\
\hline
$\sigma_S$ & 33.93 & 29.24 & 25.36 \\
$\sigma_P$ & 2.52 & 2.18 & 1.88 \\
$\sigma_{\rm tot}$ & 36.45 & 31.42 & 27.24 \\
$\sigma_P/\sigma_S~(\%)$ & 7.43 & 7.46 & 7.41 \\
$\sigma_P/\sigma_{\rm tot}~(\%)$ & 6.91 & 6.94 & 6.90 \\
\end{tabular}
\end{ruledtabular}
\end{table}

It should be mentioned that, in the hadronization of the bottom-charmed baryons, it will be accompanied by a large nonperturbative effect, which brings a large theoretical uncertainty. Especially for the color sextet intermediate diquark, there is no clear correlation with the wave functions. According to the one-gluon-exchange interaction in the diquark state, the contribution of the color sextet shall be suppressed at least by $v^2$ than the color antitriplet, or too small to be ignored \cite{Ma:2003zk, Zheng:2015ixa}, and the total events of the $\Xi_{bc}$ baryon are decreased accordingly. We therefore use $h_{\mathbf 6}=0$ as the conservative baseline and regard $h_{\mathbf 6}=v^2h_{\bar{\mathbf 3}}$ and $h_{\mathbf 6}=h_{\bar{\mathbf 3}}$ as phenomenological hadronization benchmarks. Within the NRQCD framework, the average relative velocity squared $v^2$ of the constituent heavy quarks is a crucial parameter determining the validity of the nonrelativistic expansion. Generally, the velocity scaling rules for heavy quarkonium mesons yield $\langle v^{2}\rangle_{c\overline{c}}\approx0.3$ and $\langle v^{2}\rangle_{b\overline{b}}\approx0.1$~\cite{Bodwin:1994jh,QuarkoniumWorkingGroup:2004kpm,Brambilla:2010cs}. The $[bc]$-diquark system in the color antitriplet state is more loosely bound than heavy quarkonium mesons, leading to a smaller relative velocity. To perform a conservative estimation of the theoretical uncertainty, $\langle v^2 \rangle_{bc}$ is set within the envelope range of $[0.1, 0.3]$. It is indicated that  $\mathcal{O}(v^{2})$ relativistic corrections may introduce a systematic uncertainty of 10\%--30\%. As presented in Table~\ref{csuncer}, although the different phenomenological treatments of the color sextet $h_\mathbf{6}$ shift the hadronization benchmark, the total predicted events of $\Xi_{bc}$ are enhanced by up to 61.60\%, 71.79\%, and 77.60\% for $\sqrt{s}=250$, $500$, and $1000\mathrm{~GeV}$, respectively, when $h_\mathbf{6}=h_{\bar{\mathbf{3}}}$. Finally,  it is worth emphasizing that the radial wave functions at the origin describing the transition probability $h_{\bar{\mathbf{3}}}$, namely, $|R_{bc}(0)|^2$ and $|R'_{bc}(0)|^2$ corresponding to $S$ and $P$ waves, are overall parameters of the cross sections. Therefore, this source of uncertainty acts merely as a global normalization factor; if more precise calculations (e.g., from lattice QCD or improved potential models) of these wave functions become available in the future, our baseline cross section predictions can be directly and easily adjusted.

\begin{table}
\small
\caption{Hadronization-benchmark dependence of the total cross sections $\sigma$ (in femtobarn) for $\Xi_{bc}$ photoproduction, obtained with $h_{\mathbf{6}}=0$, $h_{\mathbf{6}}=v^2h_{\bar{\mathbf{3}}}$ ($v^2=0.1$--$0.3$) and $h_{\mathbf{6}}=h_{\bar{\mathbf{3}}}$ at three collision energies (in GeV).}
\centering
\footnotesize
\begin{tblr}{
  cells = {c},
  cell{1}{1} = {r=2}{},
  cell{1}{2} = {c=3}{},
  cell{1}{5} = {c=3}{},
  cell{1}{8} = {c=3}{},
  vlines,
  hline{1,3-6} = {-}{},
  hline{2} = {2-10}{},
}
$\sqrt{s}$   & $h_{\mathbf{6}}=0$ &     &       & $h_{\mathbf{6}}=v^2 h_{\bar{\mathbf{3}}}$ &     &       & $h_{\mathbf{6}}=h_{\bar{\mathbf{3}}}$ &     &       \\
     & $\gamma+\gamma$  & $\gamma+g$ & Total & $\gamma+\gamma$  & $\gamma+g$ & Total & $\gamma+\gamma$  & $\gamma+g$ & Total \\
250  & 27.51  & 15.22  & 42.73  & 28.89--31.64   & 16.48--18.99   &  45.36--50.63   & 41.26 & 27.79 & 69.05      \\
500  & 13.46  &  31.42    &  44.88  & 14.13--15.48    &  33.97--39.07   &   48.10--54.55  & 20.18 & 56.92 & 77.10  \\
1000 & 5.77  &  56.42   & 62.19  &  6.06--6.64   &  60.96--70.03   & 67.02--76.67      & 8.66  & 101.79 & 110.45
\end{tblr}
\label{csuncer} 
\end{table}

\section{Summary}
\label{sec:4}

Within the factorization framework of NRQCD, the photoproduction of ground and excited $\Xi_{bc}$ are investigated at ILC and CLIC with three different collision energies $\sqrt{s}=250$, 500, and 1000 GeV, respectively. 
The direct and resolved photoproduction channels are both taken into account with intermediate $\langle bc\rangle$-diquark states, i.e., $\gamma+\gamma \rightarrow \langle bc\rangle[n]+\bar{c}+\bar{b} \rightarrow \Xi_{bc} +\bar{c}+\bar{b}$ and $\gamma+g \rightarrow \langle bc\rangle[n]+\bar{c}+\bar{b} \rightarrow \Xi_{bc} +\bar{c}+\bar{b}$. Twelve configurations of $\langle bc\rangle$-diquark states can be $(bc)[^1S_0]_{\bar{\textbf{3}}/\mathbf{6}}$, $(bc)[^3S_1]_{\bar{\textbf{3}}/\mathbf{6}}$, $(bc)[^1P_1]_{\bar{\textbf{3}}/\mathbf{6}}$, $(bc)[^3P_0]_{\bar{\textbf{3}}/\mathbf{6}}$, $(bc)[^3P_1]_{\bar{\textbf{3}}/\mathbf{6}}$, and $(bc)[^3P_2]_{\bar{\textbf{3}}/\mathbf{6}}$. For the conservative $h_{\mathbf 6}=0$ baseline, the two-channel combined $P$-wave contribution is 8.07\%, 7.86\%, and 7.56\% of that of the $S$ wave at $\sqrt{s}=250$, 500, and 1000~$\mathrm{GeV}$, respectively. Assuming that a 100\% decay probability of $P$-wave states into the ground state as an upper-bound estimation, the total produced events of ground-state $\Xi_{bc}$ from two subprocesses can reach $4.27\times10^5$, $4.49\times10^5$, and $6.22\times10^5$ at $\sqrt{s} = 250$, $500$, and $1000$ GeV, respectively, with an integrated luminosity of $10^4\mathrm{~fb^{-1}}$. Note that these yields only represent the parton-level inclusive rates without considering the detector acceptance and backgrounds.

Then the related differential distributions are presented, including $p_T$, $y$, $s_{35}$, $s_{45}$, $\cos\theta_{35}$, and $\cos\theta_{45}$ distributions. 
The differential distributions all emphasize that the contribution of the $S$ wave is dominant throughout the whole region compared with the $P$ wave of each subprocess. The transverse momentum distribution reveals that both in $S$ and $P$ waves, the photoproduction of $\Xi_{bc}$ through the resolved $\gamma+g$ channel is larger than that via photon-photon fusion in the region of small $p_T$. The overall trend of total $S$ and $P$ waves shows a downward trend, and there are significantly more events in the small-$p_T$ region. For the rapidity distribution for the photoproduction, the curve of $\gamma+g$ has a steeper slope, while that of $\gamma+\gamma$ is more gradual. It can be seen from the angular distributions $\cos\theta_{35}$ and $\cos\theta_{45}$ that the bottom-charmed baryon is most likely to move parallel to the two heavy antiquarks. The same conclusion has also been evident in the distributions of the invariant masses $s_{35}$ and $s_{45}$.

Finally, the heavy-quark-mass and common-scale dependences have been quantified. The choices $h_{\mathbf 6}=0$, $v^2h_{\bar{\mathbf 3}}$ ($v^2=0.1$--$0.3$), and $h_{\bar{\mathbf 3}}$ are presented as hadronization benchmarks. The quoted resolved-channel results are GRS-based central predictions, and different photon-PDF parametrizations may introduce an additional systematic dependence. The cross section of each intermediate diquark state has a strong dependence on the heavy quark masses $m_c$ and $m_b$, and it shows a monotonically decreasing trend as the heavy quark mass increases.
Regarding the photoproduction channel $\gamma+\gamma$ ($\gamma+g$), the total cross section $\sigma_{tot}$ can vary by $-30.24\% \sim+54.68\%$ 
($-36.86\% \sim+73.23\%$) for $m_c = 1.8 \pm 0.3$ GeV and  $-11.74\% \sim +13.74\%$  ($-17.06\% \sim +21.96\%$)  for $m_b = 5.1 \pm 0.3$ GeV. The contribution of $P$-wave states can be $6.83\%\sim12.00\%$  ($5.70\%\sim 10.09\%$)  of that of the $S$-wave states, $6.39\%\sim10.71\%$  ($5.39\%\sim9.17\%$) of the total cross section with the uncertainty of heavy quark mass.
The theoretical uncertainty of $\sigma_{tot}$ caused by the common scale is approximately $-26.67\%\sim+46.73\%$ ($-13.30\%\sim+16.01\%$) for $\gamma+\gamma$ ($\gamma+g$) with $\mu \in [\mu_0/2,2\mu_0]$. However, the contribution of the $P$-wave states relative to that of the $S$-wave states and the total cross section has a relatively stable proportion. Relative to the conservative $h_{\mathbf 6}=0$ baseline, the extreme benchmark $h_{\mathbf 6}=h_{\bar{\mathbf 3}}$ increases the total rate by 61.60\%, 71.79\%, and 77.60\% at $\sqrt{s}=250$, 500, and 1000~$\mathrm{GeV}$, respectively. The overall order of magnitude of the predicted inclusive rates remains unchanged.

\acknowledgments
This work was supported by Guangxi Young Elite Scientist Sponsorship Program (No. GXYESS2025005). This work was also supported by the Natural Science Foundation of Guangxi (No. 2025GXNSFAA069775) and by the National Natural Science Foundation of China (Grant No. 12505106). H.-H. M. is also supported by the S\~ao Paulo Research Foundation (FAPESP), Brazil. Process No. 2025/01276-7.

\section*{Data availability}
The corresponding results or data are already provided in our manuscript, and they can be obtained by using the parameters and relevant tools given in our manuscript.

\clearpage

\end{document}